\documentclass[11pt]{article}
\usepackage{graphicx}
\begin{document} 

\title{Competition of languages in the presence of a barrier}

\author{Christian Schulze and Dietrich Stauffer\\
Institute for Theoretical Physics, Cologne University\\D-50923 K\"oln, Euroland}

\maketitle
\centerline{e-mail: stauffer@thp.uni-koeln.de}

\bigskip
\small{Abstract: Using the Schulze model for Monte Carlo simulations
  of language competition, we include a barrier between the top half
  and the bottom half of the lattice. We check under which conditions 
  two different languages evolve as dominating in the two halves.
}


\bigskip

\section{Introduction}
Languages are influenced by natural barriers \cite{nichols}, like mountains 
\cite{nasidze}, water or politics 
\cite{schnael}. On different sides of a mountain ridge different dialects or
languages may be spoken, and the same separation happened on the two
sides of the English Channel \cite{forster}. Our previous attempt \cite{cise}
to simulate this effect with the Viviane model \cite{viviane} of language
competition \cite{abrams,chachacha,newbook,tuncay} was unsuccessful. Thus we now
try to use the Schulze model \cite{schulze,cise} to check under which conditions
one barrier leads to the domination of two different languages on the two 
sides of the barrier.

In the next section we define the model, section III gives some of our results,
and section IV summarises our work.

\section{Model}

Each site of a $L \times L$ square lattice is occupied by one adult person
speaking one language (dialect, grammar, ...). Each such language is defined
by $F$ different features each of which can take one of $Q$ different values.
Thus we have in total $Q^F$ different possible languages. We use $Q = 3$ and 5
and $F = 8$ and 16. Changes in the languages are ruled by two probabilities
$p$ and $q$. At each time step (one sweep through the lattice) each language
can change into another one by changing each of its $F$ features independently
with probability $p$. This change means that with probability $1-q$ a random 
value between 1 and $Q$ is selected, while with probability $q$ we accept
the corresponding language element from one of the four neighbours, randomly
selected. Thus $q$ means linguistic diffusion, while $p$ corresponds to 
biologial mutations.

Also, in contrast to biology, humans shift away from small to large languages
in order to be able to communicate better. Thus with probability $(1-x)^2$ at 
each iteration each person selects the whole language of one randomly selected
lattice neighbour and gives up the old language. Here $x$ is the fraction of
others speaking the same language as the old language. Normally this fraction 
was counted for the whole population, but now we calculate it from the four
nearest neighbours. We assume a horizontal barrier separating the upper half of 
the lattice from its lower half. Then we disallow this shift to another
language if that language comes from the other side of the barrier, except that
with a low crossing probability $c$ we may shift also to a language spoken
on the other side.  (The above transfer $q$ is always allowed also from the 
other side.) 

Since now we calculate $x$ from the neighbourhood only and not from the whole
population we have checked that again, as with earlier versions 
\cite{schulze,cise}, for small $q$ and large $p$ the population fragments
into numerous languages even if we start with everybody speaking the same 
language. And starting from a fragmented population we get dominance of 
one language, spoken by more than half of the population, if we use small
$p$, large $q$ and not too large $L$. 

\section{Results}

Without any barrier, Fig.1 shows how ``mutations" destroy the initial order
if we start with everybody speaking the same language. Thus Fig.2 later will
use a low probability $p = 0.05$ and a high $q = 0.9$ to facilitate the
emergence of dominating languages when initially everybody selects randomly
a language. In Fig.1 we start on the left with a small $p$ and then increase
$p$ in steps of 0.01. The observation time is $t = 1000$, and every 100 time
steps the fraction of people speaking the most widespread language is shown in 
Fig.1. Thus we see how for low $p$ this order parameter, the largest fraction,
stabilises to a value slightly below one, while for larger $p$ it decays to 
about zero. (For clarity Fig.1 is presented as one curve for different $p$, as 
if we would have started the simulation of a new $p$ with the final population
of the previous $p$. Actually, we started for each $p$ with everybody speaking
the same language. Thus in the top part of Fig.1 the first plateau corresponds 
to $p = 0.10$, the second to 0.11, the third to 0.12, followed by a decay at 
0.13.)

\begin{figure}[hbt]
\begin{center}
\includegraphics[angle=-90,scale=0.3]{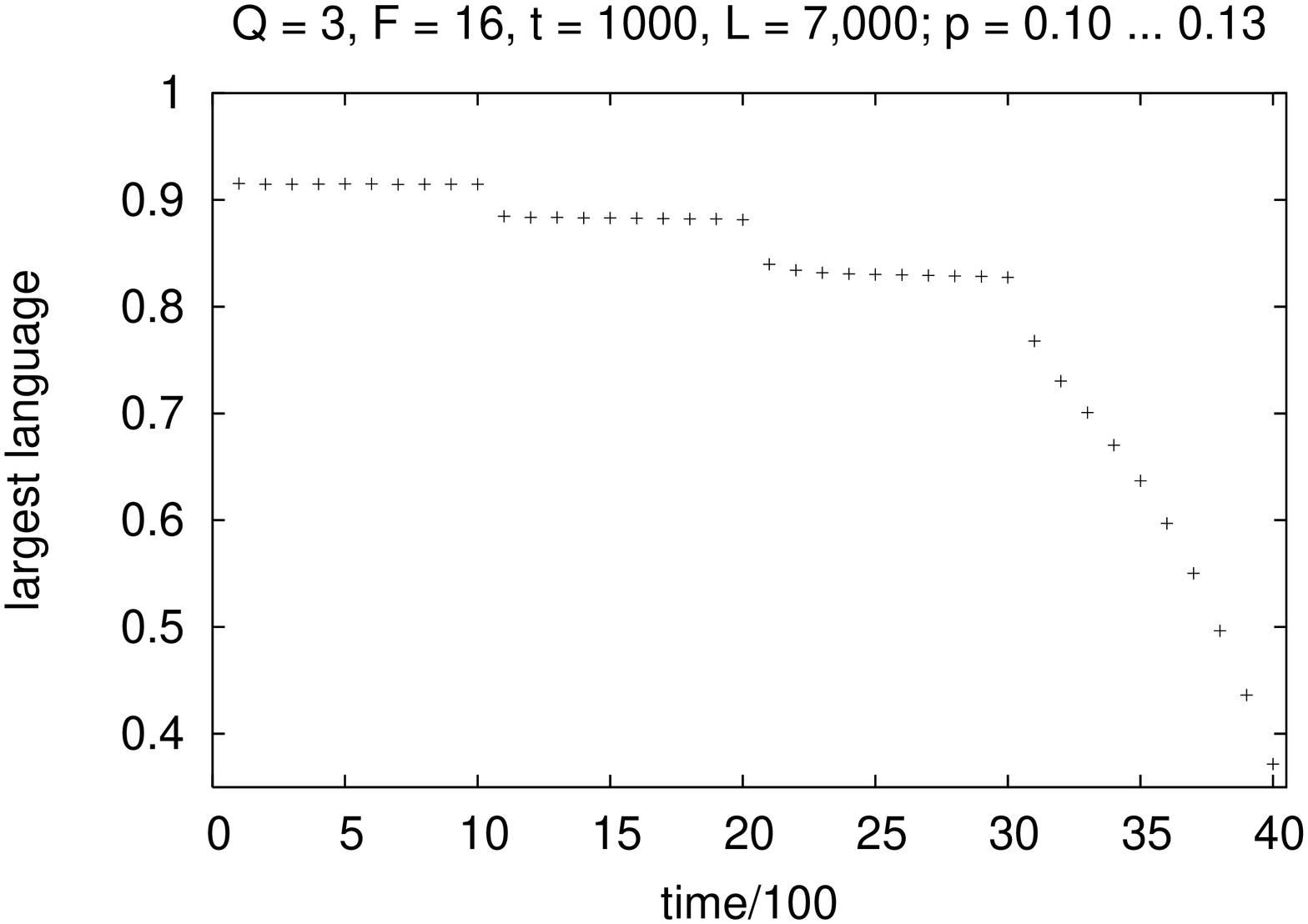}
\includegraphics[angle=-90,scale=0.3]{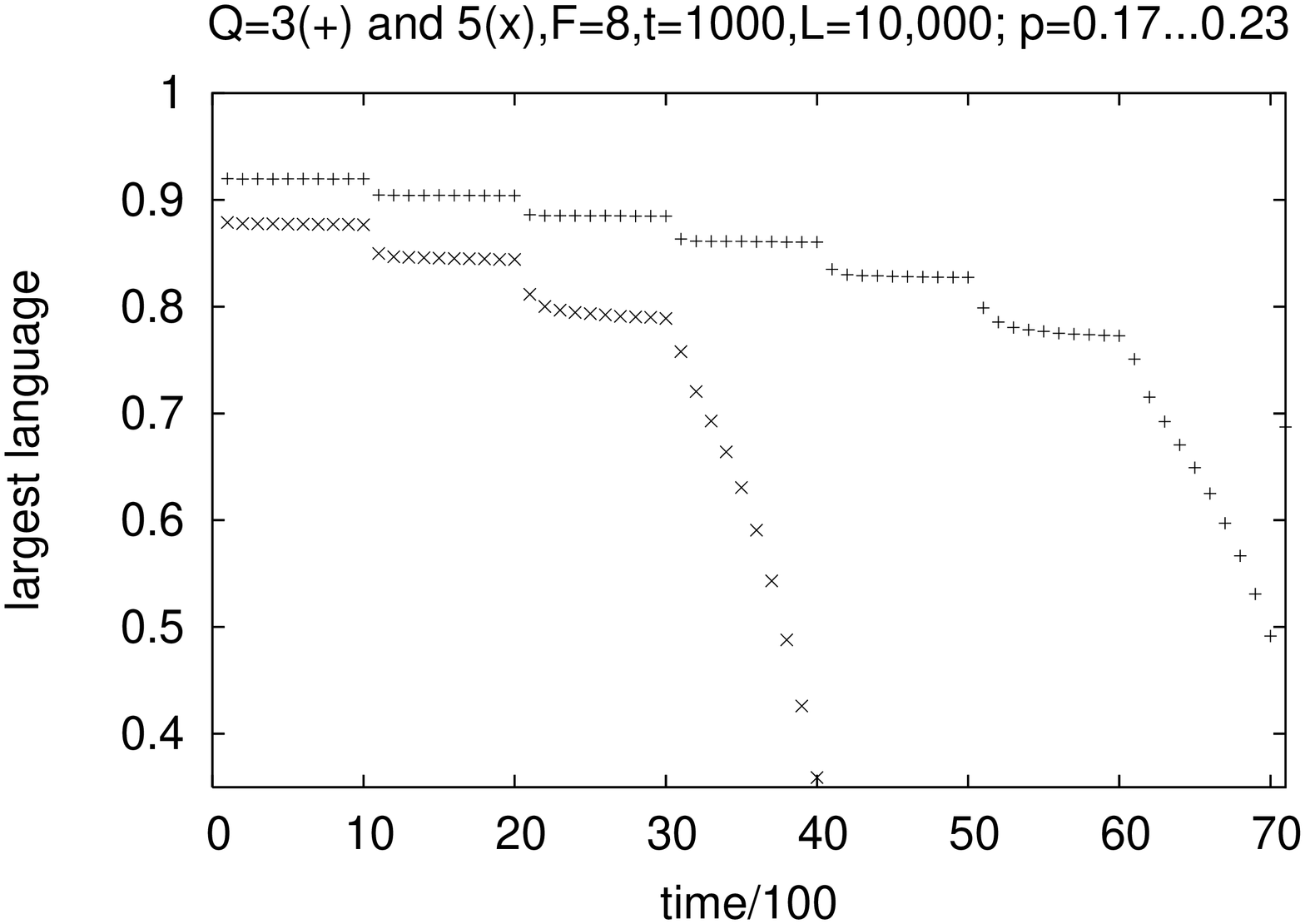}
\end{center}
\caption{Order parameter = fraction of people speaking the most widespread 
language, starting with everybody speaking the same language. Top: $Q=3,
\; F=16$; bottom: $Q=3$ and $5$, $F=8$. $p$ varies from left to right.
}
\end{figure}

\begin{figure}[hbt]
\begin{center}
\includegraphics[angle=-90,scale=0.5]{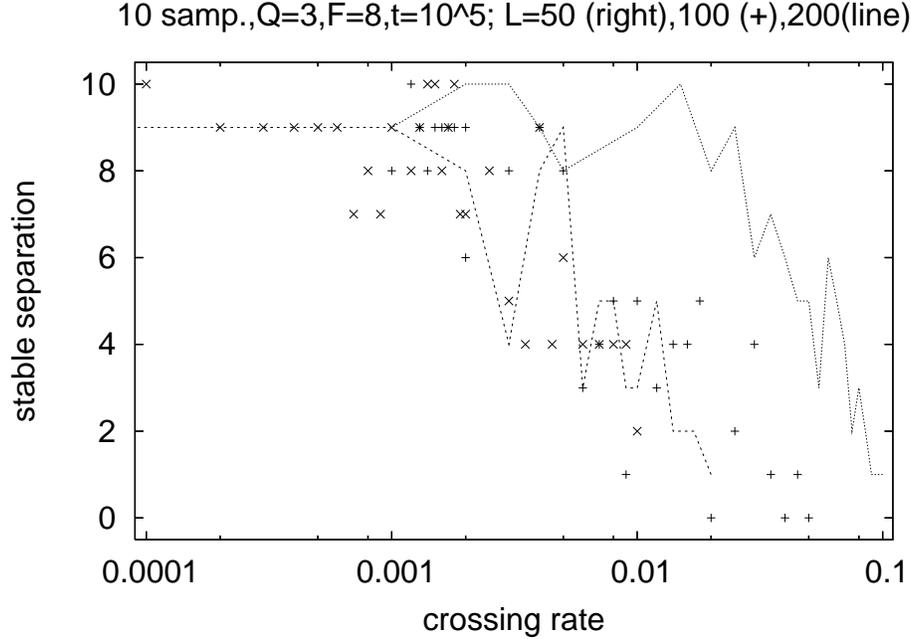}
\end{center}
\caption{Transition from stable to unstable language separation with increasing
crossing probability $c$, for $p=0.05, \; q=0.9$ and the other parameters as 
shown in the headline. The x symbols refer to $Q=5$ instead of 3, at
$L = 100$. For $F=16$ instead of 8, the transition is near $c \simeq 0.006$ 
at $L = 100$ (not shown). Stability is very rare for larger crossing rates
than shown here.
}
\end{figure}

\begin{figure}[hbt]
\begin{center}
\includegraphics[angle=-90,scale=0.5]{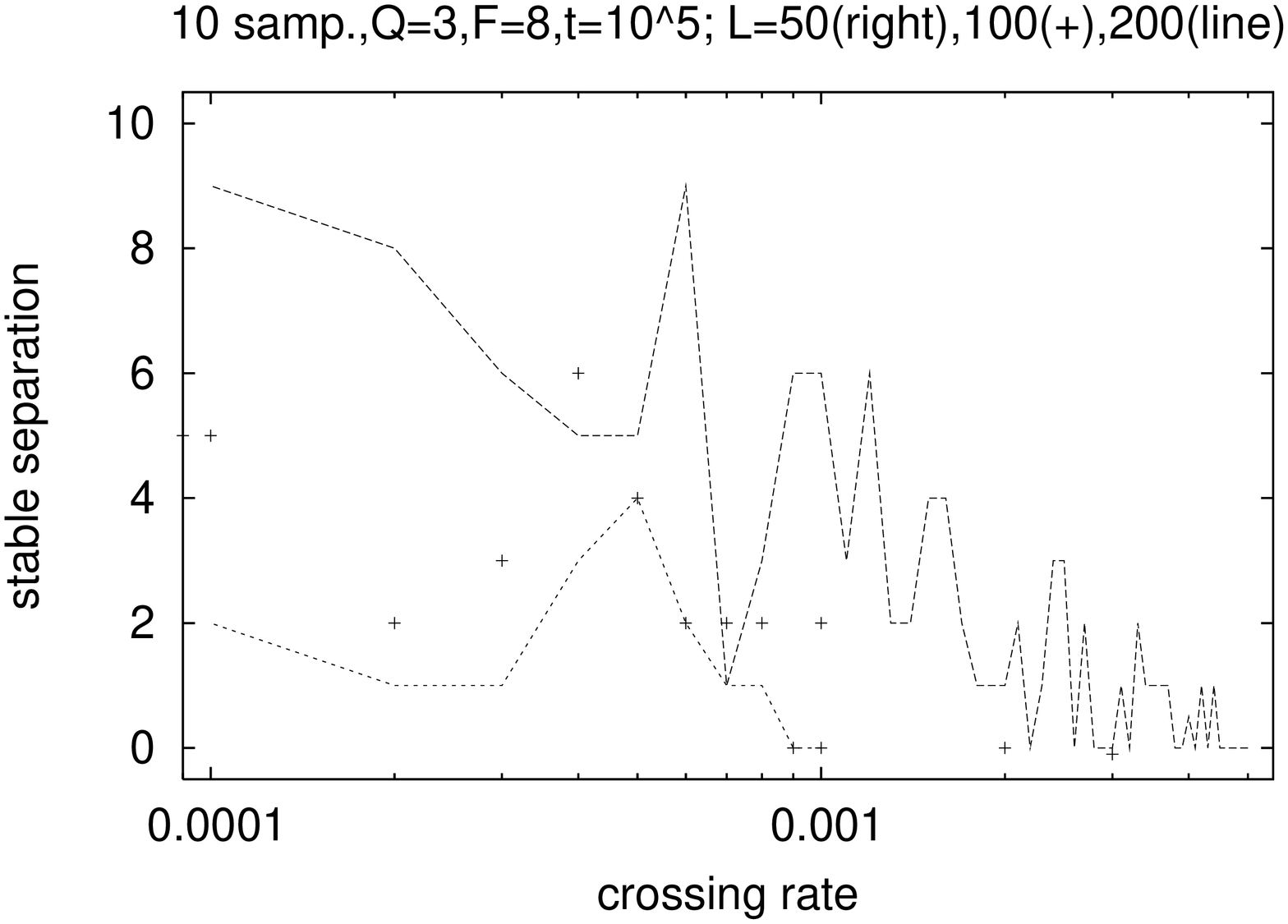}
\end{center}
\caption{As Fig.1 but for $q = 0.7$ instead of 0.9. 
}
\end{figure}

Now we include the barrier which can be crossed with a low probability $c$.
We call the situation stable if starting from random fragmentation, the most
widespread language is spoken at the end of our observation time $t$
by nearly half the population; then usually 
another language is spoken by most of the other half. Due to the coupling 
between the two lattice parts, arising from the probabilities $q$ and $c$, it 
may also happen, that after some time the same language dominates in both parts 
of the lattice; this case we call unstable since we are interested in the
coexistence of two languages, each dominating in its half of the lattice. 

It may happen that for the same set of probabilities, some random numbers give
stable and some unstable language distributions. Thus we look at ten samples
and reach the transition point
when five samples are stable and five are unstable. Fig.2
shows the transitions: Number of stable samples among the ten simulated
samples. We see a rather broad transition where that number decreases from 
(nearly) ten to (nearly) 0. And small lattices ($L = 50$) differ strongly from
larger ones ($L = 100$). Unfortunately, our changes to pure local 
interactions require long 
observation times near $10^5$ since for shorter times the order parameter
(fraction of people speaking the largest language) may not yet have grown 
sufficiently. Thus, our lattices in Fig.2 are much smaller than in Fig.1. 

Finally, Fig.3 shows our results for $q = 0.7$ instead of 0.9. Now we are 
closer to the case where ordering is impossible (the fragmented population
remains fragmented for $q < 0.42$) even at $c=0$. Thus the results are less
clear  but still show that the transitions are at much smaller $c$ than in 
Fig.2.

\section{Summary}

For low enough crossing probabilities $c$ we found stability of one language 
dominating on one side of the barrier and another language dominating on the
other side, in a variant of the Schulze model. Earlier, we were unable to get
such a seemingly trivial result in the Viviane model \cite{viviane,cise}.
The Tuncay models contain no geography. Since we are not aware of other
models for the competition of thousands of languages, our model is the first
known to us which allows for the stability of two different languages on 
different sides of a barrier. One now could apply this method to islands, i.e. 
to sections of the lattice surrounded by barriers on all four sides \cite{w}.
Since Fig.2 shows clear size effects, we then expect the transitions  
to happen at probabilities $c$ which are smaller for larger islands. 

We thank S. Wichmann, E. Holman and M. Ausloos for helpful comments.

\end{document}